# Sector-Based Radio Resource Allocation (SBRRA) Algorithm for Better Quality of Service and Experience in Device-to-Device (D2D) Communication

Pimmy Gandotra, *Student Member, IEEE*, Rakesh Kumar Jha, Senior *Member, IEEE*, Sanjeev Jain

*Abstract*—The mounting content sharing among users has resulted in a considerable rise in wireless data traffic, pressurizing the cellular networks to undergo a suitable upheaval. A competent technology of the fifth-generation networks (5G) for efficiently supporting proximity-based applications is Device-to-Device (D2D) communication, underlaying cellular networks. Significant advances have been made till date, for allocating resources to D2D users in cellular networks, such that sharing of spectral resources between cellular and D2D users is carried out in a coordinated manner. In this paper, a sector-based radio resource allocation (SBRRA) algorithm for resource block allocation to D2D pairs has been proposed, where the number of resource blocks (RBs) is allocated to each D2D pair in an adaptive manner, based on the demanded application by each pair. Different applications demand a varying number of RBs, in accordance with their priority. This algorithm focusses on the use of sectored antennas at the base station, for a better performance and low complexity. Extensive simulations are carried out, considering real-time scenario, for ensuring satisfactory Quality of Service (QoS) and Quality of Experience (QoE) by the users. The efficiency of the proposed scheme is proved by comparing it with the RB allocation using Hidden Markov Model (HMM).

## I. INTRODUCTION

An exceptional augment for mobile broadband services has been observed in the past decade due to the rise in the number of communicating devices [1] and demand for high data rate applications. With the drastic rise in the count of devices, the macro base station (BS) will have to support thousands of devices incorporated with multiple radio access technologies (RATs), at the same time. To support all these applications and meet the demands of the subscribers, prodigious research is ongoing on the 5G networks [2], which promise 1000x energy efficiency and data rate [3], with device-to-device (D2D) communication being identified as a competent technology [4], allowing direct information exchange between proximate users.

D2D communication can effectively offload traffic from the BS, a critical field identified by 3GPP [5], adjusting the power level in cellular systems [6], thus improving the network performance. Even for cells present in different tiers, D2D can serve as an effective load balancing technique [7]. It provides an adequate platform for a disruptive networking architecture, which is completely device-centric. The various benefits of D2D communication include surged energy efficiency, spectrum efficiency, extended coverage, reduced latency and support for green communication [8]. However, to utilize the benefits of D2D, the related challenges like resource allocation, power control, interference management [9], security [10] efficaciously need to be dealt with.

The hostility of the wireless communication channel and radio resource insufficiency require competent resource allocation. The problem of resource sharing in D2D communication is being addressed in this paper, using cell sectorization. In a sectorized cell, the BS is equipped with a number of directional antennas, splitting the cellular coverage area into multiple sectors. Sectorization in CDMA networks has gained significant attention in the past [11]. Better capacity [12] and spectral efficiency (SE) can be achieved through sectorization, as a result of reduced interference, without rise in capital and operational expenditures. A tri-sector architecture provides better link quality and call blocking performance. For improving the network throughput, a tri-sector cellular network is studied in [13], using stochastic geometry, which effectively balances the load and optimizes the network power. Another sector architecture is proposed in [39].

In some cases, the sectors may be overlapping, imposing interference constraints on the users within the sectors, and restraining the system performance. Even with a high degree of overlapping between the sectors, the QoS demands of the subscribers can be appropriately met [14]. Sector based resource allocation strategies also play a key role in relay networks as well [15]. Use of sector antennas also improves the secrecy of the system, as is evaluated in [16]. Apprehending the benefits of sectorization in cellular networks, this paper proposes a novel approach for adaptive RB allocation to D2D users in a 5G WCN, with the base station equipped with sector antennas.

Pimmy Gandotra is a PhD Research Scholar in Department of Electronics and Communication Engineering, Shri Mata Vaishno Devi University, J&K, India. (E-mail: pimmy.gandotra@gmail.com).
Rakesh Kumar Jha, is Assistant Professor in Department of Electronics and Communication Engineering, Shri Mata Vaishno Devi University, J&K, India. (E-mail: jharakesh.45@gmail.com).
Sanjeev Jain is Professor and Vice Chancellor, Shri Mata Vaishno Devi University, Katra.
(E-mail:dr_sanjeevjain@yahoo.com)

TABLE I
COMPARISON OF SBRRA WITH EXISTING RESOURCE ALLOCATION SCHEMES

| Reference | Objective | Modelling Technique Used | QoS requirement consideration | QoE requirement consideration | Utility Function | UL/DL |
|---|---|---|---|---|---|---|
| [22] | Addressing the resource allocation problem by cascading channel and power allocation problems | Graph theory and game theory | ✓ | ✗ | Transmission rate | DL |
| [27] | Study of resource allocation problem with Cognitive radio and D2D communication technologies | Geometric water-filling | ✗ | ✗ | Transmission rate | DL |
| [35] | To jointly study mode selection, resource allocation and interference management in D2D communication networks | Learning framework using Markov chain | ✗ | ✗ | Sum rate | DL |
| [24] | To study the resource allocation problem in multi-cast D2D | Generalized bender decomposition method | ✓ | ✗ | Sum rate | UL |
| [23] | To analyze the network performance of a D2D communication network | Stochastic geometry | ✗ | ✗ | Coverage probability | UL |
| [25] | To study a distributed resource allocation technique, with transmit power minimization | Q-learning | ✗ | ✗ | Total transmission power | UL |
| [34] | To enable multiple D2D transmissions in sectored cell | Power-Emission based modelling | ✓ | ✗ | Throughput | UL |
| SBRRA | To adaptively allocate optimal number of resource blocks to all the demanding D2D pairs in a sectored cell | Sector-based heuristic scheme, using HMM | ✓ | ✓ | Throughput, MOS | DL |

*A. Related Work*

The number of cellular users, D2D users, channel gains, the signal-to-interference-plus-noise ratio (SINR) is highly uncertain in a wireless communication channel, due to its random nature. These aspects must be disparagingly addressed, before radio resource allocation (RRA). Prior to resource allocation, selection of mode for the user equipment (UE) is essential. In [18], the decision on the selection of mode (cellular/D2D), is made on the basis of the received signal strength threshold value.

Initial research on resource sharing between cellular and D2D users considered single RB reuse case, where a single D2D pair can reuse the RBs of a single cellular user only. Though this is a simple consideration, sufficient resource efficiency is not guaranteed in such networks, drifting the research towards multi-RB reuse scenarios [20].

Resource sharing between cellular and D2D users in a cellular network for different modes is studied in [21]. The resource allocation problem in [22] is fragmented in to channel allocation and power control problems, which are then solved using graph theory and game theory, respectively. The authors in [23] propose a stochastic geometry based framework for resource allocation in D2D networks. For a multi-cast D2D scenario, resource allocation problem has been addressed in [24]. A resource allocation scheme, in conjunction with transmit power minimization has been discussed in [25], and is based on Q-learning [26]. Resource allocation in a cognitive radio (CR) D2D network has been evaluated in [27], introducing an adaptive subcarrier allocation scheme, followed by the use of geometric waterfilling technique [28] for power allocation.

With video applications gaining a significant stimulus in 5G networks, QoS guarantees becomes paramount, with throughput serving as the key performance indicator. The overall experience of the end user is characterized by the QoE, indicated by the MOS. A three-step resource allocation strategy for maximizing the network throughput has been proposed in [29], which guarantees the QoS requirements of the cellular users as well as D2D users, along with optimal power control. Another resource allocation scheme for guaranteeing the quality of service (QoS) requirements of the users has been proposed in [30], using the column generation method. It improves the spectrum efficiency of the networks significantly. QoS enhancement in [42] is achieved through buffer-size limitation of the cellular users.

A QoE aware resource and power allocation scheme has been presented in [31], where the Mean Opinion Score (MOS) of the users has been quantified. An algorithm for QoE enhancement using D2D communication has been proposed in [32]. Significant gains in the performance has been achieved with the proposed scheme. QoE driven schemes [33] can act as driving factors for D2D communication in the approaching era of wireless networks.

Most of the existing work on resource allocation in D2D communication consider omni-directional antennas deployed at the BS. As the wireless communication is heading towards 5G, an era of UDNs is anticipated. In such high-density networks, the count of users, and that too D2D users will be very large, resulting in profound interference problem. This problem can be effectively dealt, with the use of sectored antennas at the BS [17]. This will boost the throughput, as is contemplated in [34], [19]. This provokes emphasis on the use of sector-partitioned cellular regions. Although existing algorithms have been efficient in management and reduction of interference [35], [36], incorporating sector antennas in unison with these will have a positive impact on the system performance. A comparison of the existing techniques with SBRRA has been given in Table I.

*B. Contribution:*

In this paper, a sector based radio resource allocation (SBRRA) scheme is proposed, for D2D communication in a tri-sectored 5G WCN, considerably reducing interference and meeting the QoS and QoE requisites of the users with low complexity. The proposed scheme follows an adaptive RB allocation mechanism, where an appropriate RB count is allocated to each demanding D2D pair, in accordance with the demanded applications. Three applications are assumed in this paper: Non-Conversational Video (A1), Conversational Video (A2) and conversational voice (A3), in decreasing order of their priority [38]. The main contributions of this paper embrace the following:

- Under a random deployment of users, the proximate users in each cell sector, adhering the distance constraint, form D2D pairs, which share RBs with their cellular reuse partners. These reuse partners are identified, on the basis of the channel gains between the cellular and D2D users. The D2D pair count varies iteratively, thus, their cellular reuse partners are also iteratively updated.
- The optimization problem for throughput maximization is formulated, for guaranteed QoS and QoE in the considered network. Since the proposed SBRRA scheme facilitates RB reuse in successive iterations, a substantial augmentation in throughput is achieved.
- Simulation results obtained with the proposed scheme are compared with adaptive RB allocation using HMM, with the proposed SBRRA scheme outperforming HMM, in terms of throughput and MOS. Sector antenna reduces interference, and thus complexity, which has been analyzed through simulations.

The remaining paper is organized as follow. The system model and problem formulation is given in Section II. Resource Allocation using HMM is discussed in Section III. Flowcharts and pseudo codes are contained in Section IV and Section V illustrates SBRRA scheme through an example. The simulation results are analyzed in Section VI. The paper finally concludes in Section VII.

## II. SYSTEM MODEL AND PROBLEM FORMULATION

The system model for investigating RB allocation in a D2D network is introduced in this section, followed by the problem formulation.

*A. System Model*

A cellular coverage area of radius '$\dot{R}$' meters is considered, with a tri-sector BS at the center (Fig. 1). Since a D2D communication scenario is considered, RB sharing between cellular and D2D users result in interference.

Users within each sector are assumed to be randomly distributed. The total number of users in each sector is assumed to be equal, and denoted as N. As a result of random user distribution, during every iteration, variable number of D2D pairs will be formed per sector, depending on the distance between the users (Algorithm 1). The set of D2D pairs in a sector is denoted by $\check{D} = \{1, 2, \dots d\}$, and the set of cellular users is denoted by $\mathbb{C} = \{1, 2, 3, \dots c\}$. Each cellular user is periodically allocated equal count of RBs by the BS, denoted by the set $\mathbb{R} = \{1, 2, \dots r\}$. The channel state information (CSI) is assumed to be known at the BS, along with the location of the D2D users, which is obtained through the global positioning system (GPS). As a result, no overhead is involved. The values of *d* and *c* are dynamically updated during each iteration, illustrating a real-time scenario (Fig. 1). The RB structure is also shown in Fig. 1.

Transmission powers of the BS ($P_B$), and users are assumed to be fixed. The D2D transmitter power is evenly distributed over all the RBs. At each iteration, every D2D pair in $\check{D}$ needs to determine the cellular user from which it will share the RBs. This is determined from the existent channel conditions between the D2D pairs and cellular users. The cellular user possessing the highest channel gain with a D2D pair will serve as its RB sharing partner. Depending on the demanded application, different pairs share different number of RBs from their respective sharing cellular partners. The demanded applications for all the pairs are determined in an adaptive manner, using HMM. The set of applications is denoted by $\boldsymbol{Ap} = \{A1, A2, A3\}$.

The SINR at the receiver of $j^{th}$ D2D pair, sharing $k^{th}$ RB of $i^{th}$ cellular user, $j \in \check{D}$, $k \in \mathbb{R}$, and $i \in \mathbb{C}$, for every iteration is given by

$$SINR_j^{k,i} = \frac{P_j^{k(i)} h_j^{k(i)}}{\sigma_{N0} + \mathbb{I}_j} \tag{1}$$

where $\mathbb{I}_j$ denotes the total interference encountered by $j^{th}$ D2D pair and is given by

$$\mathbb{I}_j = \mathbb{I}_{B(j)} + \mathbb{I}_{i(j)} + \mathbb{I}_{j'(j)}$$
$$= P_B h_{B,j}^{k(i)} + \sum_{\substack{k' \in \mathbb{R} \\ k' \neq k}} P_i^{k'(i)} h_{i,j}^{k'(i)} + \sum_{\substack{j' \in \check{D} \\ j' \neq j}} P_{j'}^{k(i)} h_{j,j'}^{k(i)} \tag{2}$$

Multiple factors contribute to interference in a downlink communication scenario. The BS keeps continuously transmitting and coordinating the cellular and D2D users. Also, cellular and D2D communication are going on concurrently, causing the active pair to suffer interference from the BS, denoting $\mathbb{I}_{B(j)}$ as

$$\mathbb{I}_{B(j)} = P_B h_{B,j}^{k(i)} \tag{3}$$

Each pair shares a certain count of RBs from its cellular partner, which is application dependent. The RBs remaining with the cellular user cause interference to the D2D transmissions, due to their own simultaneous transmissions over the remaining RBs. The interference due to the remaining RBs with the $i^{th}$ cellular user, to the $j^{th}$ pair is represented as

$$\mathbb{I}_{i(j)} = \sum_{\substack{k' \in \mathbb{R} \\ k' \neq k}} P_i^{k'(i)} h_{i,j}^{k'(i)} \tag{4}$$

Since we are talking of an ultra-dense deployment in the 5G networks, which will contain a massive number of D2D pairs close to each other, the pairs within a certain range of every other pair, sharing the same RBs will result in co-tier interference. If $j'$ is the number of pairs in proximity to pair *j*, interference due to these pairs is

$$\mathbb{I}_{j'(j)} = \sum_{\substack{j' \in \check{D} \\ j' \neq j}} P_{j'}^k h_{j,j'}^k \tag{5}$$

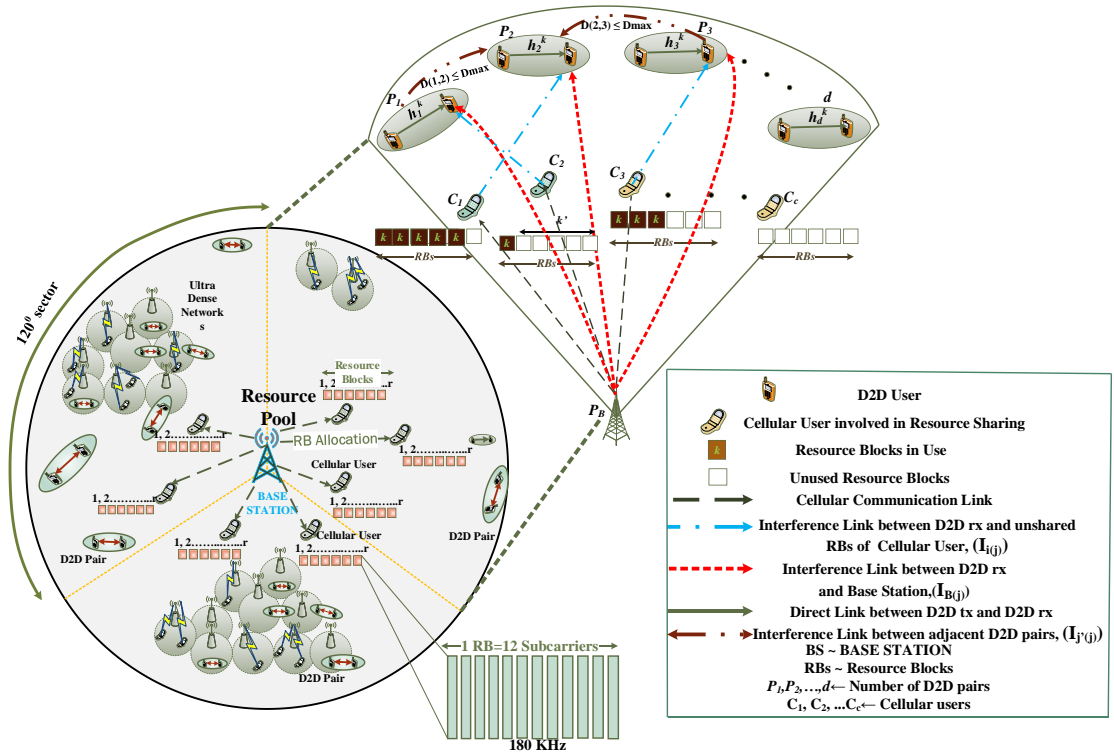

Fig. 1. system model

This interference factor comes into effect only if distance between the adjacent pairs is less than or equal to $D_{max}$. Summation of (3), (4) and (5) gives the resultant total interference. The symbols used are given in Table II.

Using the Shannon capacity formula [40], the achievable throughput can be computed for the $j^{th}$ pair, for iteration '$n$' as,

$$T_j^i(n) = \beta \sum_{k \in \mathbb{R}} \log_2(1 + SINR_j^{k,i}) \quad (6)$$

where $\beta$ is the bandwidth of one RB.

For $d$ pairs in each iteration, the total throughput for a particular iteration is the summation of throughputs of $d$ pairs, i.e.

$$T_d(n) = \sum_{\substack{j \in \check{D} \\ i \in \mathbb{C}}} T_j^i(n) \quad (7)$$

Multiple iterations are considered in our scenario, with the total number of iterations denoted by the set $\mathbb{Q} = \{1, 2, \ldots q\}$. If a cellular user '$i$' is involved in RB sharing in multiple iterations, with different D2D pairs, then the resultant throughputs of the D2D pairs are successively added. When $i^{th}$ cellular user shares RBs with $l$ pairs, for $n$ number of iterations, then the throughput achieved $l$ pairs, such that $l \in \check{D}$, is given by

$$T_l^i = \sum_{n \in \mathbb{Q}} \sum_{p=1}^{l} \sum_{k \in \mathbb{R}} \log_2(1 + SINR_p^{k,i}) \quad (8)$$

Here, $p$ denotes the $p^{th}$ D2D pair at $n^{th}$ iteration, sharing RBs of $i^{th}$ cellular user. (Refer **Lemma 1**) ∎

*Proposition 1:* The $i^{th}$ cellular user is involved in RB sharing during successive iterations, depending on its channel state with the formed pairs, and if it has at least $r/2$ RBs remaining with it, after sharing in a previous iteration. This is assumed to be a necessary condition for sustaining D2D and cellular communication simultaneously for the chosen applications.

The equation (8) shows that when same cellular user is used many times for RB sharing, the resultant throughput is boosted. This process of aggregation of throughputs continues till the time the cellular user remains active in RB sharing, at different iterations. Thus, the total throughput of the system is the average throughput for all iterations, denoted by

$$T_{system} = \frac{\sum_{n \in \mathbb{Q}} T_d(n)}{\sum n} \quad (9)$$

TABLE II
LIST OF SYMBOLS USED

| Symbol | Description |
|---|---|
| $r$ | Total number of resource blocks with each cellular user |
| $Ap$ | Set of applications |
| $d$ | Total number of pairs |
| $c$ | Total number of cellular users |
| $q$ | Total number of iterations |
| $SINR_j^{k,i}$ | SINR for the receiver of $j^{th}$ D2D pair, sharing $k^{th}$ RB of $i^{th}$ cellular user |
| $P_j^{k(i)}$ | Transmission power of $j^{th}$ D2D transmitter on $k^{th}$ RB of $i^{th}$ cellular user |
| $h_j^{k(i)}$ | Channel gain between D2D transmitter and D2D receiver, for $j^{th}$ pair, sharing $k^{th}$ RB of $i^{th}$ cellular user |
| $\sigma_{N0}$ | Receiver Noise Power |
| $D_{max}$ | Maximum Distance between neighboring D2D pairs, which can cause interference |
| $SINR_j^{threshold}$ | Threshold SINR for pair $j$ |
| $P_j^{maximum}$ | Maximum transmission power for pair $j$ |
| $h_{B,j}^{k(i)}$ | Interference channel gain from the BS to receiver of $j^{th}$ D2D pair, sharing $k^{th}$ RB of $i^{th}$ cellular user |
| $P_i^{k'(i)}$ | Transmission power of $i^{th}$ cellular user on its RB $k'$ |
| $h_{i,j}^{k'(i)}$ | Channel gain of interference between unshared RBs ($k'$) of $i^{th}$ cellular user and transmitter of $j^{th}$ D2D pair |
| $P_{j'}^{k(i)}$ | Transmission power of D2D transmitters other than pair $j$, causing interference and sharing $k^{th}$ RB of $i^{th}$ cellular user |
| $h_{j,j'}^{k(i)}$ | Channel gain of interference between pairs $j$ and $j'$ |

After the throughput computation, for ensuring QoE to all the users in the network, MOS is calculated. For $j^{th}$ pair, during $n^{th}$ iteration, the MOS is given by [33]

$$MOS^j(n) = 5 - \frac{578}{1 + \left(\frac{T_j^i(n) + 541.1}{45.98}\right)^2} \quad (10)$$

where $T_j^i(n)$ is the throughput expressed in Kbps.

*Proposition 2:* The MOS is measured on a five-point scale, with the values ranging between 1 to 5. A value of '5' of the MOS signifies excellent quality, and a value '1' denotes least quality. Values 2 and 3 are annoying quality signals. A general satisfaction to the end user can be assured through a value between 4 to 4.5, as these provide a good quality signal.

### B. Problem Formulation

Our primary goal is the maximization of the overall throughput, for ensuring the QoS to all the demanding D2D pairs in the network. The optimization problem is formulated as

$$\max_{\mathbb{Q}, \check{\mathbb{D}}, \mathbb{C}} T_{system} \quad (11)$$

s.t.

$$SINR_j^{k,i} \geq SINR_j^{threshold} \quad (11(a))$$
$$\sum_{k \in \mathbb{R}} P_j^k \leq P_j^{maximum} \quad (11(b))$$
$$1 \leq k(i) \leq (r-1) \quad (11(c))$$
And, $\quad r - k(i) \geq \frac{r}{2} \quad (11(d))$

Constraint (11(a)) ensures sufficient number of RBs available to each D2D pair, as needed by the demanded application, thus meeting the minimum QoS requirement and (11(b)) ensures that the total transmission power of a D2D transmitter over all RBs cannot exceed its maximum transmission power. The number of shared RBs is one less than the total available, as in (11(c)), so that at least one RB is available to the cellular user for simultaneous transmission with D2D pairs. A particular cellular user, providing favorable channel gain in successive iterations to the formed pairs also needs to have sufficient number of remaining RBs $(r - k(i))$ to be able to share RBs again, which is ensured by (11(d)). $T_{system}$ depends on the number of pairs formed at any instant, the number of cellular users involved in RB sharing to the pairs, the number of RBs shared between the two types of users in the network, and the number of iterations over which the throughput is computed.

### III. RESOURCE ALLOCATION USING HMM

RB allocation from cellular users to D2D users can be performed using HMM. Adequate number of RBs are allocated to the D2D pairs, on the basis of the demanded application, as stated earlier. BS controls the scheduling of users in the network. The complete HMM process is represented by a set of states, S and a set of parameters, ʉ [37]. The states of the model are represented using the state diagram, as shown in Fig. 2, with the three states: base station (BS), cellular user and the D2D pair, and represented as S= (Base station, Cellular User, Pair), in reference to Table III.

TABLE III
PROBABILITIES FOR ADAPTIVE RESOURCE BLOCK ALLOCATION USING HMM

| Demand | Response | | |
|---|---|---|---|
| | Base Station | Cellular User | D2D Pair |
| Base Station | 0.02 | 0.8 | 0.18 |
| Cellular User | 0.19 | 0.01 | 0.8 |
| D2D Pair | 0.18 | 0.8 | 0.02 |

When once, the states, S, have been decided, parameters are represented by the probability matrices л, A, B, such that, ʉ = {л, A, B}, where the first states of a sequence are given by the prior probabilities, л; transition from one state to another is given by the transition probabilities, *A* and the likelihood of an observation is represented by emission probabilities, *B*. Additionally, an HMM operation is characterized by two sequences, namely the hidden state sequence, $Q = \{q_1, q_2, \ldots q_N\}$ and the observation sequence, $X = \{x_1, x_2, \ldots x_N\}$. The probability of a hidden state sequence is given by the product of transition probabilities

$$P(Q|ʉ) = л_{q1} \prod_{n=1}^{N-1} a_{qn}, a_{qn+1} \quad (12)$$

Having knowledge of the previous observation sequence, the likelihood of an observation sequences $X = \{x_1, x_2, \ldots x_N\}$, and set of parameters ʉ are prediction based and is the product of emission probabilities, given by

$$P(X|Q,,ʉ) = \prod_{n=1}^{N} P(x_n|q_n, ʉ) \quad (13)$$

The likelihood of the observation sequence is computed using trellis diagram. For a particular state sequence, knowing the previous observation sequence, likelihood can be predicted by multiplying observation and transition likelihoods along a particular path in the trellis diagram.

Equal count of RBs are allocated by the BS to all the cellular users. These are the observations, known prior the RB allocation process. Data is trained on the basis of the known observation sequence. The set of demanded applications, *Ap* and their priorities are known. The initial choice of parameters is very critical in HMM and thus, should be carefully chosen. HMM performs RB allocation to the D2D pairs formed during each iteration within the network, based on the data trained, and the set of probabilities computed.

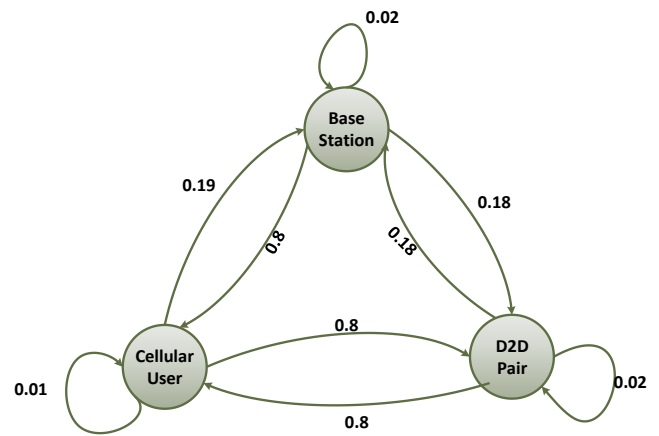

Fig. 2. Probability Distribution Diagram for adaptive RB allocation, using HMM

## IV. REALIZATION AND REPRESENTATION OF SBRRA

This section presents the pseudo code for providing a useful insight to implementation of RB allocation in D2D communication underlaying cellular networks. The pseudo codes are divided into three parts, with Pseudo code 1 explaining the realization of the scenario for D2D communication, adaptive RB allocation using HMM explained in Pseudo Code 2, and the implementation of SBRRA scheme in Pseudo code 3. For understanding the pseudo code in a better way, flowchart representations have been provided, as illustrated in Fig. 3, 4 and 5.

For 'N' number of randomly distributed users in a sector of a cellular network, the process of D2D pair formation is depicted in Fig. 3 and in Pseudo Code 1. Pair formation between a set of users is distance-dependent. When the distance constraint is fulfilled, a valid D2D pair is formed otherwise, cellular communication continues. Accordingly, the number of pairs and cellular users are updated in sets $\check{D}$ and $\mathbb{C}$, respectively.

**Pseudo Code 1 for realization of scenario for D2D communication underlaying cellular networks**

***Step 1: Input Parameters***
    Base Station: $P_B$
    Cellular User: $P_i$, $\mathbb{R}$, $r$
    D2D user: $P_j^{maximum}$
    Channel: N, $\sigma_{N0}$, $\dot{R}$, $\mathbb{Q}$, $A$
***Step 2: Initialization*:**
    Generate random user locations within
    the sector of radius $\dot{R}$, **for Iteration** *n*
    Initialize number of cellular users, $c = 0$
    /*All the cellular user indices are contained in set $\mathbb{C}$*/
    Initialize the number of D2D pairs, $d=0$
    /* All the D2D pair indices are contained in the set $\check{D}$ */
***Step 3: Check users forming D2D pairs (Conditional statements and Loops)***
    **for** *x*=1: N
    **for** *y*=1: N
    Compute distance, with the following Eq.

$$d(x,y) = \sqrt[2]{\left(X_{loc(x)} - X_{loc(y)}\right)^2 + \left(Y\_loc(x) - Y\_loc(y)\right)^2}$$

    **if** $d(x,y) \leq d_0$
    **if** $d(x,y) \neq 0$     **then**
    $\check{D} \leftarrow (x,y)$/*pair formation; Pair updated in set $\check{D}$*/
    Eliminate user *x* and *y* from the count N; N=N-2
    **end if**
    **end if**
    **end for**
    **end for**
    Rest of the users updated in set $\mathbb{C}$
    $c \leftarrow N$    /* Number of cellular users are equal to N */

**Pseudo Code 2 for resource block allocation using HMM**

***Step 1: Data training and calculating prior probabilities***
    Compute the set of probabilities and train the data
***Step 2: Specifying Applications***
    Specify the applications, in descending order of priority,
    $A1 > A2 > A3$
***Step 3: Decide number of RBs to allocate***
    RB allocation performed adaptively, in accordance with the priority of the application demanded. The value of '$k$' is application-dependent
***Step 4: SINR computation based on adaptive RB allocation***
    Obtain the SINR values from the probability distribution and trained data
***Step 5: Throughput computation***
    Compute throughput for each pair (from obtained SINR)

An adaptive RB allocation scheme using HMM is depicted in Fig. 4, allocating RBs on the basis of the demanded application. The priorities of the applications are set beforehand, determining the required RB count for each application. Depending upon the trained data, and the set of probabilities, predictions of the next states are made stochastically, and RBs are allocated to the pairs formed in the network. The sequence of steps is briefed in Pseudo code 2.

After formation of pairs, their channel states with the cellular users are determined in SBRRA. Highest channel gain providing cellular user are chosen for RB allocation, and the count of RBs is application dependent, determined with HMM. The RB allocation process with SBRRA, is given in Fig. 6, and Pseudo Code 3. After completion of RB allocation, thereafter, throughput and MOS values are computed. A cellular user can repeatedly share RBs with the pairs, during different iterations, only if it has sufficient number of RBs available with it, after sharing in the previous iteration.

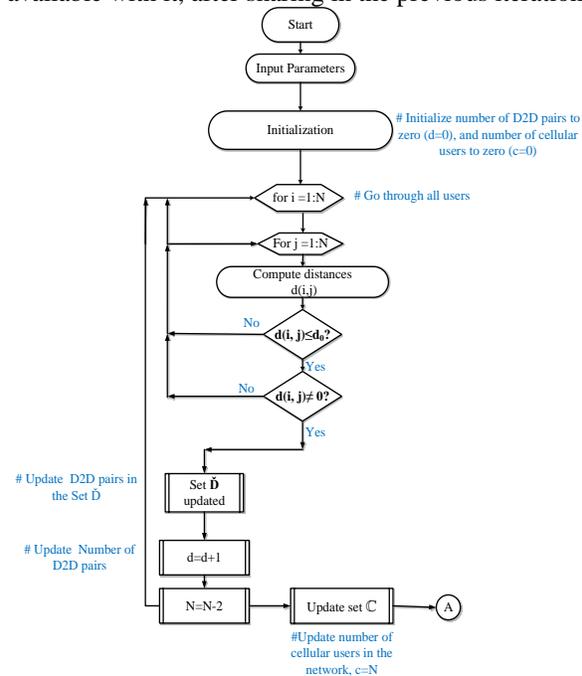

Fig. 3. Flowchart for pair formation

**Pseudo Code 3 for the proposed resource block allocation scheme (SBRRA)**

***Step 1: Determine the channel conditions for initiating RB allocation***
    Loop statement

**for** *j*=1: *d*    /* Go through all pairs and cellular users to
**for** *i*=1: *c*     obtain the channel states for Iteration n*/
Compute path loss and channel gains.
Obtain indices cellular users, Index_n (*j.i*)
**end for**
**end for**

**Step 2: Sorting the channel gains**
    Sort computed gains and indices (Index_n (*j, i*)) in descending order.  /* Index of cellular user providing best channel gain is obtained */
Obtain the indices of cellular users, allocating RBs to the demanding pairs, Index_n (*j*, 1)

**Step 3: Go through all pairs after obtaining cellular partners**
    **for** *j*=1:
    Identify the demanded application  /* This determines the value of k i.e. number of RBs shared with $i^{th}$ cellular user (From Pseudo Code 2)*/
    Allocate $k(i)$ RBs to $j^{th}$ pair
    Compute the number of RBs remaining with $i^{th}$ cellular user, i.e. $r-k(i)$
    Compute $SINR_j^{k,i}$ and then, $T_j^i$
    Compute MOS for $j^{th}$ pair
    **end for**
    Go to Step 1 and repeat for the next iteration (*n+1*), to obtain  Index_n+1(*j,i*)  /*Due to random deployment, cellular user indices and gains are updated iteratively*/

**Step 4: Check for RB reuse condition**
    *(Loop statements)*
    **for** *j*=1: *d*
    **for** *i*=1: *c*
    **if** Index_n + 1(*j, i*) == Index_n(*j, i*) /*same CUEs*/
    **if** $r - k(i) \geq r/2$  **then**
    Repeat Step 3
    Total Throughput will be the sum of throughput obtained in Step 3 and Step 4
    Compute $MOS^j$
    **else**
    Cellular user in Index_n+1 (*j,i*) is used for RB  sharing with $j^{th}$ pair in Iteration *n+1*
    Repeat Step 3, and Compute $MOS^j$
    **end if**
    **end if**
    **end for**
    **end for**
    Repeat the above steps for *q* number of iterations
**Step 6: Plot the results**

### A. Computational Complexity in SBRRA

Since SBRRA is an iterative scheme for adaptive RB allocation, the channel conditions are repeatedly checked between the cellular users and D2D pairs. These help in carrying out the transmission under the best channel states. The main idea of SBRRA is to reuse the RBs of cellular users which have been active during the previous iterations. Such a scheme supports better QoS and QoE, as is depicted through simulation results.

Considering a 5G scenario, with UDNs, the problem of interference management is paramount. Use of sector

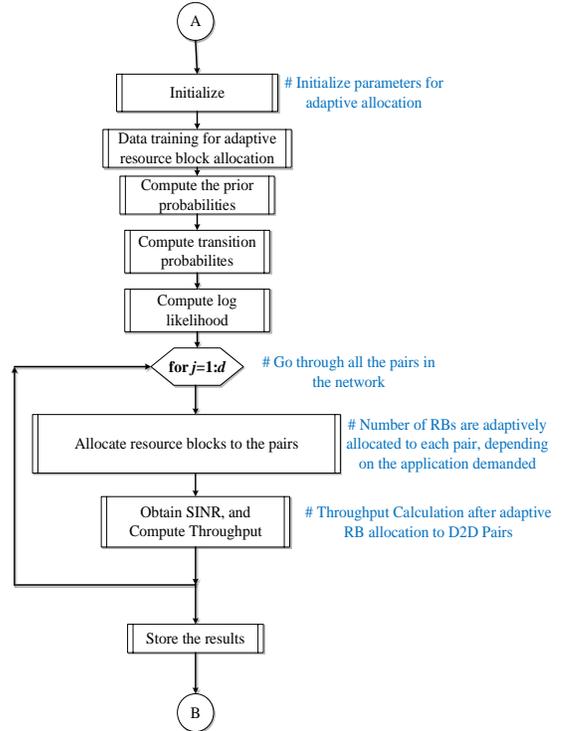

Fig. 4. Flowchart for adaptive RB allocation using HMM

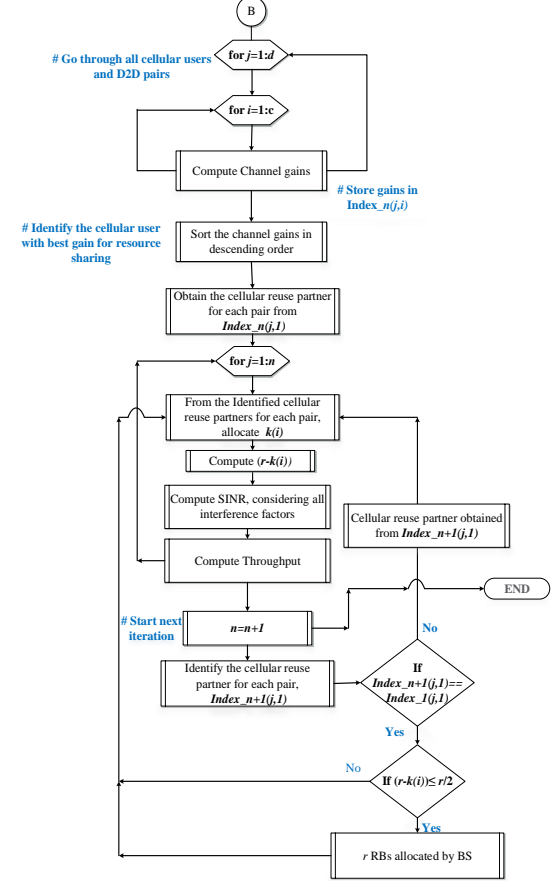

Fig. 5. Resource block Allocation using SBRRA algorithm

approach is beneficial in interference reduction, as has been stated in Section I. The pairs formed will suffer interference due to multiple factors (Eq. (2)). This interference, $\mathbb{I}_j$ rises with an upsurge in pair count, but its overall impact is less This article has been accepted for publication in a future issue of this journal, but has not been fully edited. Content may change prior to final publication. Citation information: DOI 10.1109/TVT.2017.2787767, IEEE Transactions on Vehicular Technology

8with the sector approach in a cell, than without the use of sector approach. This aspect is considered in SBRRA, for analyzing the complexity of the system. This paper, however, considers the worst-case complexity as well. If the pairs formed have their locations close to each other, being concentrated in a region within a sector, then $\mathbb{I}_j$ rises due to higher $\mathbb{I}_{j'(j)}$. At such instants, rise in complexity is inevitable. Such a complexity is very genuine for a realistic scenario.

## V. SBRRA: AN ILLUSTRATIVE EXAMPLE

The entire process of RB allocation, using the proposed SBRRA strategy, has been depicted in Fig. 6. For a random deployment of users in a single sector, a case for *d*=2 is assumed, each pair demanding a distinct application in each iteration, from set ***Ap***. The number of cellular users in the sector are assumed to be five (*c*=5), with $\mathbb{R}$ RBs allocated by the BS, in each iteration. Here, the active cellular users are those which allocate RBs to the demanding pairs. In Iteration 1, cellular user 1 ($CU_1$) is considered as the active cellular user in the network, having favorable channel conditions (providing the highest gain) with both the pairs ($Pair_1$ and $Pair_2$). Let us assume that $Pair_1$ is demanding, application A1 ($k(i) = 5$), while $Pair_2$ is demanding A3 ($k(i) = 1$), and these are allocated the demanded RB count by $CU_1$. The throughput for $Pair_1$ and $Pair_2$, in Iteration 1 is $T_1^1(1)$ and $T_2^1(1)$, respectively (From Eq.(6)). The total throughput of the system in Iteration 1 is

$$T_{total}(1) = T_1^1(1) + T_2^1(1) \qquad (14)$$

To be able to allocate RBs again to the pairs, in another iteration, and for continued information exchange over its RBs, $CU_1$ will be requiring sufficient RBs. In the middle of the duration of sharing of RBs with pairs, $CU_1$ may have to operate in cellular mode as well. This will also need RBs. Thus, it is allocated '*r*' RBs again, by the BS, in the beginning of the next iteration, if the RBs remaining with it are less than the threshold (*r/2*).

Checking for the channel conditions again for the two pairs, and the cellular users in the network, it is observed that the cellular users offering the highest channel gain to both the pairs vary iteratively. Accordingly, the throughputs are computed. The cellular user, not providing favorable channel conditions are assumed inactive. With reference to Fig. 6 and Table IV, the cellular partners for both pairs can be obtained. The total throughput of the system in Iteration 2 is

$$T_{total}(2) = T_1^2(2) + T_2^3(2) \qquad (15)$$

Since, after RB allocation to the respective pairs, $CU_2$ and $CU_3$ become RB deficient (in Iteration 2), they are allocated '$\mathbb{R}$' RBs by the BS, in Iteration 3, and these now have ($\mathbb{R}$+1) RBs available for sharing. In this iteration, different applications are demanded by the pairs, and have different cellular users for sharing of RBs. Such an allocation of RBs by the BS, to the cellular users in the network, assures continuity of service. Such a mechanism provides QoS, which is the main aim of the proposed scheme. Apart from QoS guarantee to the D2D pairs, the QoS of cellular users is also simultaneously promised in this method, though no mathematical analysis has been given in the paper.

A similar procedure of allocation continues for the successive iterations. In Iteration 3, the cellular reuse partners for $Pair_1$ and $Pair_2$ are $CU_4$ and $CU_2$, respectively, sharing different number of RBs, as per demanded application. Consequently, throughputs are $T_1^4(3)$ and $T_2^2(3)$, for $Pair_1$ and $Pair_2$, respectively. In this iteration, the RBs of $CU_2$ are shared for the second time. Thus, throughputs of D2D pairs, achieved by sharing RBs of $CU_2$ in the second iteration are added and the total throughput of the system in Iteration 3 is

$$T_{total}(3) = T_1^4(3) + (T_2^2(3) + T_2^2(2)) \qquad (16)$$

TABLE IV
AN ILLUSTRATION OF RESOURCE BLOCK ALLOCATION PROCESS FOR *d*=2

| Iteration No. | Cellular User involved in RB sharing | Application demanded by $Pair_1$ and throughputs | | Application demanded by $Pair_2$ and throughputs | |
|---|---|---|---|---|---|
| | | $App^n$ | Throughput | $App^n$ | Throughput |
| 1. | $CU_1$ | A1 | $T_1^1(1)$ | A3; | $T_2^1(1)$ |
| 2. | $CU_2, CU_3$ | A1 | $T_1^2(2)$ | A1; | $T_2^3(2)$ |
| 3. | $CU_4, CU_2$ | A3 | $T_1^4(3)$ | A3; | $T_2^2(3)$ + $T_2^2(2)$ |
| 4. | $CU_5, CU_4$ | A2 | $T_1^5(4)$ | A3; | $T_2^4(4)$ + $T_1^4(1)$ |
| 5. | $CU_5, CU_4$ | A2 | $T_1^5(5)$ + $T_1^5(4)$ | A3 | $T_2^4(5)$ + $T_2^4(4)$ + $T_1^4(3)$ |

Similarly, throughputs for $Pair_1$ and $Pair_2$ in Iteration 4 and Iteration 5 are computed, reusing RBs of previously active cellular users, with the total system throughput in Iteration 4 given by

$$T_{total}(4) = T_1^5(4) + (T_2^4(4) + T_1^4(3)) \qquad (17)$$

and for Iteration 5 given as

$$T_{total}(5) = \underbrace{(T_1^5(5) + T_1^5(4))}_{\text{For } Pair_1} + \underbrace{(T_2^4(5) + T_2^4(4) + T_1^4(3))}_{\text{For } Pair_2} \qquad (18)$$

Addition of throughputs is in accordance with the proposed SBRRA scheme (Eq. (8)), which promotes the sharing of RBs with cellular users that are providing good channel states in successive iterations. The total system of the throughput is obtained as an average over the different iterations, as given in Eq. (9), i.e.

$$T_{system} = \frac{T_{total}(1) + T_{total}(2) + T_{total}(3) + T_{total}(4) + T_{total}(5)}{5} \qquad (19)$$

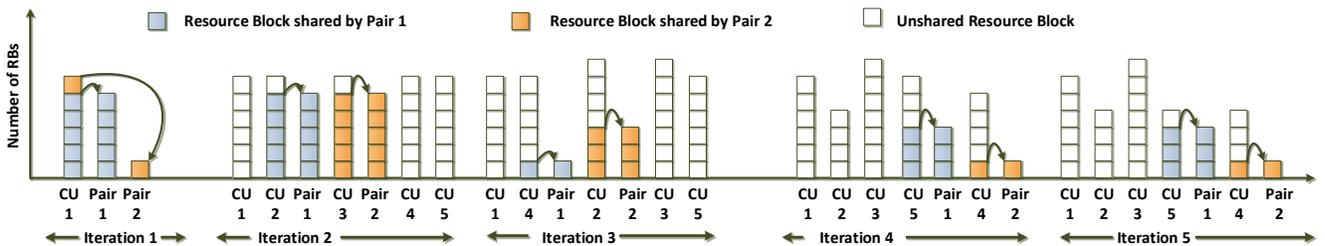

Fig. 6. Illustration of Resource Block allocation process for *d*=2, and *c*=5

0018-9545 (c) 2017 IEEE. Personal use is permitted, but republication/redistribution requires IEEE permission. See http://www.ieee.org/publications_standards/publications/rights/index.html for more information.

The number of RBs shared by the two pairs is shown in different colors, for easy understanding of the readers. The values of $k(i)$ are assumed for explanation. For a high priority demanded application, more number of RBs are allocated by the favorable cellular user in the network, and thus, higher is the throughput achieved (Eq.(6)).

## VI. SIMULATIONS AND DISCUSSIONS

The simulation parameters for RB allocation to D2D pairs have been illustrated in Table V. The channels among the users are modeled as path loss channel models [41]. Within a cell, with the BS at the cell center, use of sectored antennas is considered, dividing the cell into three $120^0$ sectors. For a realistic scenario, the value of N is highly variable. Assuming a low, medium and high density of users, simulations are conducted for N=30, 50 and 100, respectively. For a random deployment of users in the cell sector, different number of pairs may be formed at different instances.

In this section, the results from simulations have been elaborated, and inferences drawn have been stated, for adaptive RB allocation to D2D pairs, on the basis of the demanded application.

TABLE V
SIMULATION PARAMETERS

| Parameter | Value |
|---|---|
| Cell Radius, Ŕ | 500m |
| Carrier Frequency | 2GHz |
| Receiver Noise Power | -106 dBm |
| Threshold distance, $d_0$ | 20m |
| Transmission power of BS, $P_B$ | 43dBm |
| Maximum Transmission power of D2D transmitter, $P_j^{maximum}$ | 21dBm |
| Maximum Transmission power of cellular user | 24dBm |
| Resource Block bandwidth, $\beta$ | 180KHz |
| Total Bandwidth of System | 1.4 MHz |
| Number of RBs, $r$ | 6 |
| Number of Users per sector, N | 30, 50, 100 |
| Antenna Pattern | Tri-sectored |
| $D_{max}$ | 50m |
| Path loss and shadow fading at a distance $d$ km, $PL_{Bu}$ (in dB) | $148.1+37.6log_{10}d$ |
| Path loss for $d \leq d_0$ (in dB) | $40log_{10}d(in\ km)+30log_{10}fc(MHz)+49$ |
| Channel Gain at a given Path Loss (PL) | $10^{(-Path\ Loss/10)}$ |

### A. Case I: For low user density (N=30)

In this case, a random deployment of 30 users in a sector is considered., resulting in a variable number of D2D pair formation. $d$=2, 3 and 4, for five different iterations are assumed, each demanding a different application, during each iteration (Table VI) and the performance indicators over these iterations are analyzed for performance evaluation.

TABLE VI
NO. OF PAIRS FORMED, AND DEMANDED APPLICATION, FOR N=30,

| N | Iteration | d | Application Demanded |
|---|---|---|---|
| 30 | 1 | 2 | A1, A2 |
|  | 2 | 2 | A1, A1 |
|  | 3 | 2 | A3, A1 |
|  | 4 | 2 | A1, A2 |
|  | 5 | 2 | A1, A3 |
| 30 | 1 | 3 | A1, A1, A3 |
|  | 2 | 3 | A1, A1, A2 |
|  | 3 | 3 | A1, A2, A1 |
|  | 4 | 3 | A2, A1, A3 |
|  | 5 | 3 | A2, A1, A2 |
| 30 | 1 | 4 | A1, A1, A1, A2 |
|  | 2 | 4 | A3, A1, A1, A1 |
|  | 3 | 4 | A1, A3, A3, A3 |
|  | 4 | 4 | A1, A3, A1, A2 |
|  | 5 | 4 | A1, A1, A1, A2 |

The reason for the use of sectored approach is depicted in Fig. 7. In the previous generations of cellular networks, cellular radius of 1km and 2km have been existent. The cell size has been shrinking progressively, enabling efficient spectrum reuse and enhanced network capacity. When using the sector approach in traditional cellular networks with large cell size, the number of D2D pairs formed is one, two and even zero, at times as shown However, with a 500m sector, the number of D2D pairs formed is up to six, with the total number of users in the network being same in each of the three cases (N=30). Maximum number of pairs are obtained with 500m sector, thereby offloading traffic efficiently in the 5G networks, though, a larger pair count in 1km and 2km sector is also possible, if the random users in these sectors adhere to the distance constraint. Area is not a constraint upon the number of pairs in the sector. Their locations, and fulfilment of the condition for pair formation is crucial.

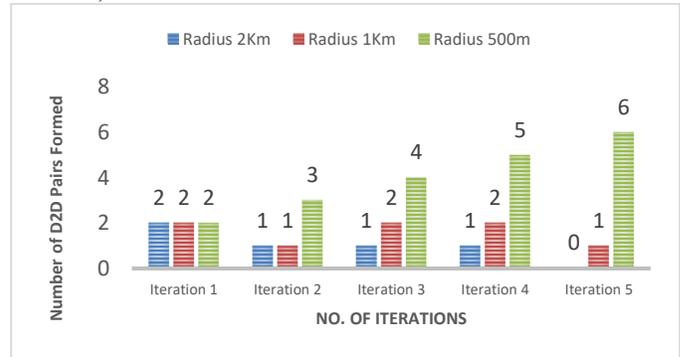

Fig. 7. Number of D2D pairs formed for 2km, 1km and 500m sector radius, for N=30

TABLE VII
SINR VALUES FOR $d$=2, 3 and 4, FOR DIFFERENT ITERATIONS

| Iterations | $d$=2 | | $d$=3 | | $d$=4 | |
|---|---|---|---|---|---|---|
|  | HMM | SBRRA | HMM | SBRRA | HMM | SBRRA |
| 1. | 23 | 25 | 31 | 50 | 48 | 77 |
| 2. | 13 | 36 | 24 | 75 | 70 | 78 |
| 3. | 17 | 56 | 45 | 71 | 46 | 45 |
| 4. | 7 | 57 | 25 | 53 | 50 | 75 |
| 5. | 16 | 43 | 26 | 58 | 58 | 96 |

SBRRA allows RB sharing between D2D pairs and cellular users on the basis of channel conditions. Higher SINR values are thus obtained, since transmissions are carried out over the best channel states. SINR v/s No. of Iterations plot is depicted in Fig. 8, for $d$=2, 3 and 4. SINR values are also given in Table VII. For each iteration, the SINR achieved with SBRRA, for every pair, is higher than HMM. The reason for the superiority of SBRRA over HMM is that SINR computation in HMM is done probabilistically, purely based on the trained data. Unlike in SBRRA, where SINR computation involves inclusion of all the interference factors in a real-time deployment scenario. This interference is reduced with the use of sector antennas, thus higher SINR values are obtained. Also, SINRs are dependent on the demanded application. Therefore, in correspondence to the listed applications in Table VI, it can be seen that greater demand of A1 corresponds to higher SINR values.

The applications demanded by all the pairs are determined adaptively, and consequently the required number of RBs are allocated to the pairs, by their cellular reuse partners. A higher priority (A1 here) application requires more number of RBs, thus, achieves higher throughput (Eq. (6)). If A1 is demanded, more number of times during different iterations, throughput achieved is higher (Eq.(8)), followed by those obtained with A2 and A3. A throughput v/s Number of Iterations plot is shown in Fig. 9, for different applications, during the different iterations. SBRRA always results in higher throughput than HMM, for the same set demanded applications by each pair, due to resource reuse from the same cellular users in SBRRA. This is in accordance with the diverse Quality of Service (QoS) requirements of the applications of the evolving wireless networks. Throughput enhancement can be less or more, depending on the application in demand, thus allocated count of RBs, during the successive iteration.

A higher throughput value also signifies a better utilization of the limited available RBs, through SBRRA since RB wastage is being avoided at each step, by making use of the RBs of same cellular user successively. The total throughput for HMM and SBRRA has been compared for different iterations, for the different number of pairs, as shown in Fig. 10. In each iteration, the throughput of SBRRA is higher than the HMM. Additionally, with increasing number of pairs, throughput of the system enhances. This is due to the reduced interference levels with the use of sectored antenna. SBRRA is thus, favorable for the 5G networks. The interference is reduced between the users lying at the sector edge and within the sector. For $d$=3, there is a fall in throughput at Iteration 4. The reason for lesser throughput can be a low-priority demanded application, or, non-reuse of RBs from the cellular users, which remained active during previous iterations. Also, for pairs formed in a concentrated region, close to each other, received SINRs are low due to high $\mathbb{I}_{j'(j)}$, resulting in low throughput.

Apart from a high throughput, to ensure the QoS, another important metric to be affirmed by the operators is the QoE. To evaluate the QoE, MOS has been computed. MOS values for different $d$, and different iterations, is shown in Table VIII. Comparing MOS values from HMM and the proposed SBRRA scheme, the value of MOS for every pair is higher with SBRRA, than in HMM, in each iteration. This assures QoE to the subscribers. The trends in MOS range is similar as that for throughput. A high QoE can be assured by the cellular network operators (values between 4 to 5), with both the schemes, a remarkably better one with the proposed SBRRA.

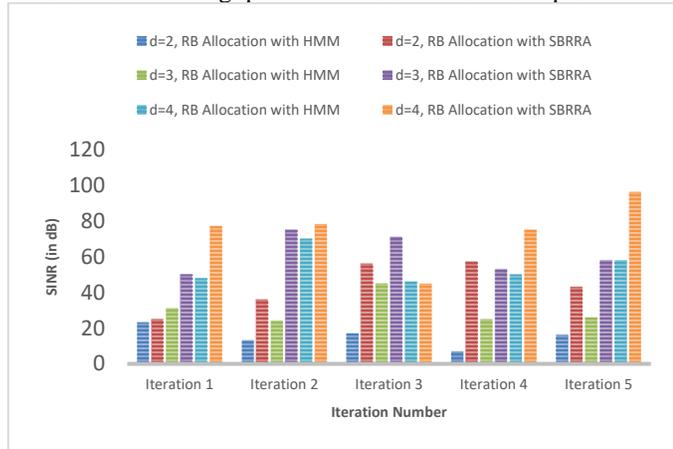

Fig. 8. SINR v/s number of iterations, for d=2, 3, 4; comparison between HMM and SBRRA, N=30

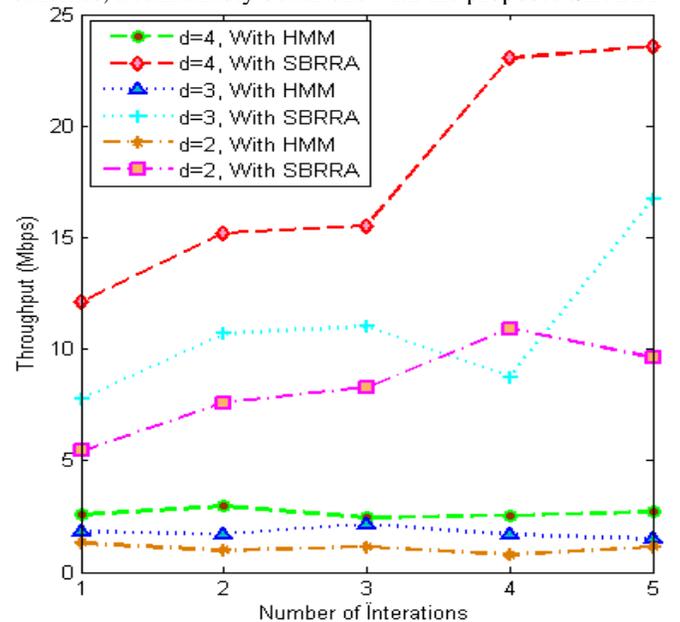

Fig. 10. Throughput v/s number of iterations for $d$=2, 3 and 4; comparison between HMM and SBRRA

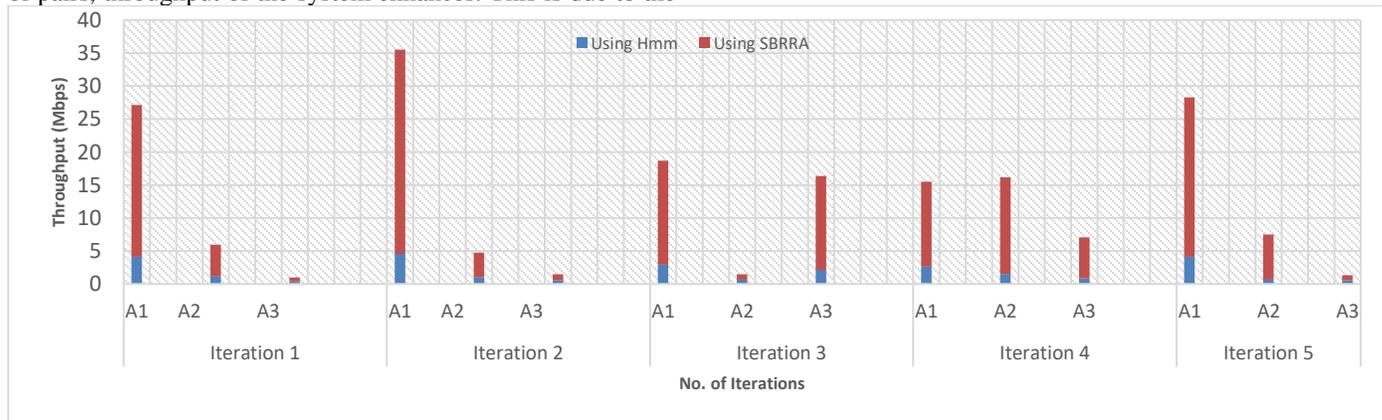

Fig. 9. Comparison of Throughput for different applications, using HMM and SBRRA

TABLE VIII
MOS VALUES FOR *d*=2, 3 and 4, FOR DIFFERENT ITERATIONS

| Iterations | d=2 | | d=3 | | d=4 | |
|---|---|---|---|---|---|---|
| | HMM | SBRRA | HMM | SBRRA | HMM | SBRRA |
| 1. | 4.14 | 4.83 | 4.016 | 4.6 | 4.112 | 4.9 |
| 2. | 3.74 | 4.91 | 3.96 | 4.87 | 4.24 | 4.81 |
| 3. | 4.025 | 4.74 | 4.22 | 4.53 | 4.02 | 4.46 |
| 4. | 3.58 | 4.91 | 3.95 | 4.82 | 4.07 | 4.9 |
| 5. | 4 | 4.64 | 3.62 | 4.83 | 4.13 | 4.91 |

### B. Case II: For medium user density (N=50)

Since 5G networks mention ultra-dense deployments, in order to study the trends in the network performance with rise in the number of users, this case evaluates the performance of the considered scenario with N=50. This count marks a drift towards higher user density. The users are again randomly deployed in the sector, with variable number of pairs formed in the sector. The number of pairs vary from 1 to 10 in this case. $d = 3, 5, 6 \text{ and } 7$ is considered in this case, each pair demanding a distinct application, for different iterations, as depicted in Table IX.

The results in Fig. 11, illustrate a comparison of the total number of pairs formed in a sector of radius 2km, 1km and 500m, with maximum number of pairs formed in the 500m, while lesser number of pairs are formed in 1km and 2km radius sectors. Maximum pair count with N=50 is higher, than N=30, though $d = 2, 3 \text{ and } 4$ exist in case of N=50 also. This shows the random nature of the system.

TABLE IX
NUMBER OF PAIRS FORMED AND APPLICATIONS DEMANDED FOR N=50

| N | Iteration | n | Application demanded |
|---|---|---|---|
| 50 | 1 | 3 | A3, A1, A3 |
| | 2 | 3 | A3, A1, A1 |
| | 3 | 3 | A2, A1, A1 |
| | 4 | 3 | A1, A3, A1 |
| | 5 | 3 | A1, A1, A1 |
| 50 | 1 | 5 | A3, A2, A1, A1, A3 |
| | 2 | 5 | A1, A2, A2, A1, A1 |
| | 3 | 5 | A1, A1, A3, A1, A1 |
| | 4 | 5 | A2, A2, A1, A2, A1 |
| | 5 | 5 | A3, A1, A1, A1, A3 |
| 50 | 1 | 6 | A1, A3, A2, A1, A3, A3 |
| | 2 | 6 | A3, A3, A1, A1, A3, A1 |
| | 3 | 6 | A1, A1, A3, A3, A2, A1 |
| | 4 | 6 | A1, A1, A1, A3, A1, A1 |
| | 5 | 6 | A1, A3, A2, A3, A1, A1 |
| 50 | 1 | 7 | A3, A2, A3, A2, A2, A1, A1 |
| | 2 | 7 | A3, A1, A2, A1, A2, A2, A1 |
| | 3 | 7 | A3, A3, A1, A1, A1, A1, A3 |
| | 4 | 7 | A3, A2, A2, A1, A3, A2, A2 |
| | 5 | 7 | A3, A2, A2, A3, A1, A1, A1 |

The SINR analysis for different number of pairs has been performed in Fig. 12, for different iterations. SINR with SBRRA is higher than in case of HMM, for most cases, for each iteration (except at some iterations and values of *d*). SINR values fall due to the co-tier interference. In this paper, the co-tier interference is considered if the distance between the adjacent pairs is less than or equal to $D_{max}$. The value of $D_{max} = 50m$ is assumed, since the maximum distance for pair formation is 20m. For a practical scenario, if we take two adjacent pairs, with 20m range, are assumed to cause interference to each other with a precision distance of $\pm 10m$ existent between them. Thus, $D_{max} = 50m$ is assumed, for an ultra-dense environment. For an iteration, the interference on a particular pair may be only due to a single adjacent pair. This adjacent pair, however, might be suffering interference from multiple other pairs in its adjacency. Thus, the interference has a cumulative impact.

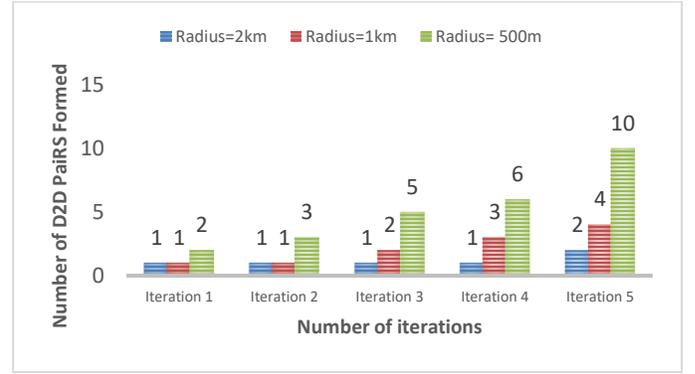

Fig. 11. Number of D2D pairs formed for 2km, 1km and 500m sector radius, for N=50

TABLE X
SINR VALUES (in dB), FOR *d*=3, 5, 6 and 7, FOR DIFFERENT ITERATIONS, WITH N=50

| Iterations→ | | 1 | 2 | 3 | 4 | 5 |
|---|---|---|---|---|---|---|
| d=3 | HMM | 31 | 38 | 29 | 17 | 32 |
| | SBRRA | 23 | 77 | 47 | 109 | 65 |
| d=5 | HMM | 40 | 83 | 100 | 93 | 95 |
| | SBRRA | 147 | 94 | 85 | 118 | 72 |
| d=6 | HMM | 102 | 121 | 128 | 91 | 114 |
| | SBRRA | 74.7 | 142 | 93 | 75 | 171 |
| d=7 | HMM | 179 | 112 | 170 | 189 | 171 |
| | SBRRA (concentrated pairs) | 121 | 92 | 144 | 101 | 125 |
| | SBRRA (distributed pairs) | 272 | 201 | 302 | 341 | 282 |

For many of the iterations and *d* values, it can be seen in Fig. 12 and Table X that SINR with SBRRA is less than with HMM. It is because for a real-time testbed, the pairs formed may be close to each other, i.e. dense in a region, within the sector, or distributed, resulting in an escalation in interference among D2D pairs, decreasing SINR values, or vice-versa. For pairs scattered over the entire sector, experiencing reduced co-tier interference, SINR values obtained are higher, and the efficacy of SBRRA is validated, in terms of RB allocation over favorable channel conditions. The variations are depicted in Fig. 13. Thus, for different number of pairs, the pair locations may be concentrated, or distributed in a region, in the sector, lowering or improving the SINR values, respectively, as shown in Fig. 13, for the case of *d*=7.

The total throughput for different applications has been analyzed, for different iterations, in Fig. 14. Again, the maximum throughput is achieved for highest priority application, i.e. A1, followed by A2 and A3, in order of their priorities

The total throughput of the system for different iterations, for different number of pairs has been analyzed in Fig. 15. The system throughput for different number of pairs is rising with each iteration, due to the reuse of RBs from the same cellular users, as far as possible. However, the maximum achievable throughput is less for *d*=7, than for *d*=6, for SBRRA. This is due to the increasing interference from the neighboring pairs, in a concentrated pair density, which result in a decrease in the SINR value (Table X), thus supporting reduced achievable throughput. In spite of reuse of RBs from the cellular users, throughput values fall, due to rise in the co-tier interference. If instead, pairs are distributed in the sector, SINR values obtained are better, resulting in higher throughput, correspondingly.

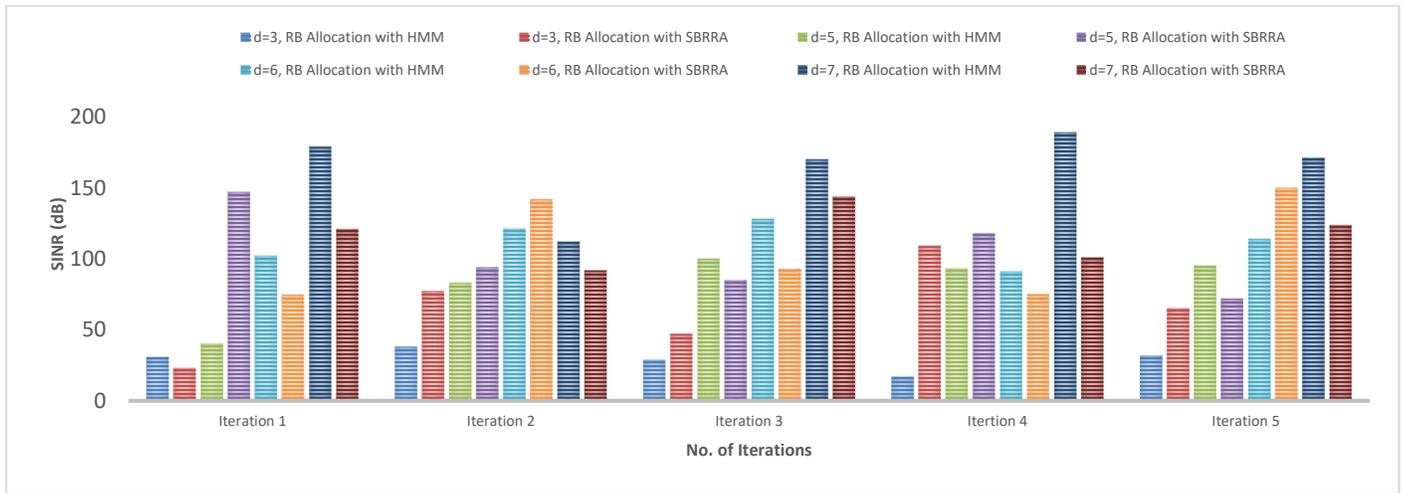

Fig. 12. SINR v/s number of iterations, for different no. of pairs formed; comparison between HMM and SBRRA, N=50

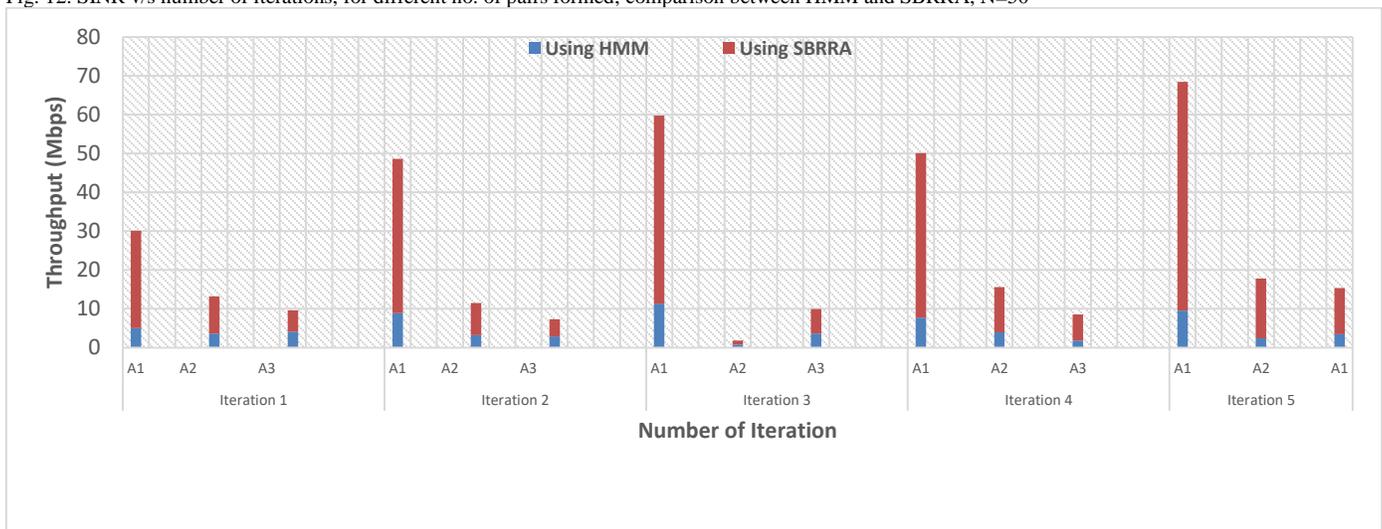

Fig. 14. Comparison of Throughput for different applications, using HMM and SBRRA

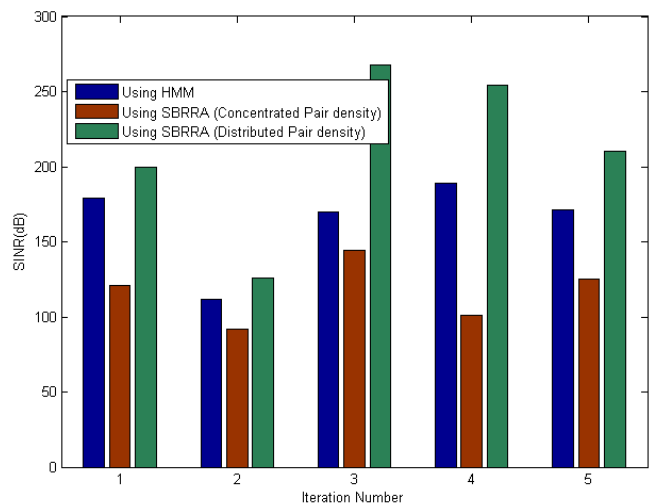

Fig. 13. Comparison of SINR obtained with HMM, SBRRA (concentrated pair density) and SBRRA (distributed pair density), for different iterations for $d=7$, N=50

Next, for investigating the QoE of the users, the MOS values have been analyzed, tabulated in Table XI. A higher MOS for SBRRA is obtained, than for the case of HMM, assuring a better QoE, even when the number of pairs formed is large. The need for an accelerated QoE, with high pair count, is successfully met with SBRRA. Even with concentrated pair density in the sector, MOS is appreciable. This depicts the competency of SBRRA, in a dense deployment.

TABLE XI
MOS VALUES FOR d=3, 5, 6 and 7, FOR DIFFERENT ITERATIONS

| Iterations→ | | 1 | 2 | 3 | 4 | 5 |
|---|---|---|---|---|---|---|
| $d=3$ | HMM | 3.88 | 4.15 | 4.03 | 3.44 | 3.7 |
| | SBRRA | 4.16 | 4.77 | 4.89 | 4.76 | 4.92 |
| $d=5$ | HMM | 3.51 | 4.21 | 4.23 | 4.25 | 4.27 |
| | SBRRA | 4.68 | 4.87 | 4.69 | 4.9 | 4.8 |
| $d=6$ | HMM | 3.9 | 3.56 | 4.28 | 3.91 | 4.21 |
| | SBRRA | 4.13 | 4.65 | 4.66 | 4.725 | 4.85 |
| $d=7$ | HMM | 4.24 | 4.19 | 4.24 | 4.21 | 4.25 |
| | SBRRA (concentrated) | 4.65 | 4.76 | 4.71 | 4.34 | 4.7 |

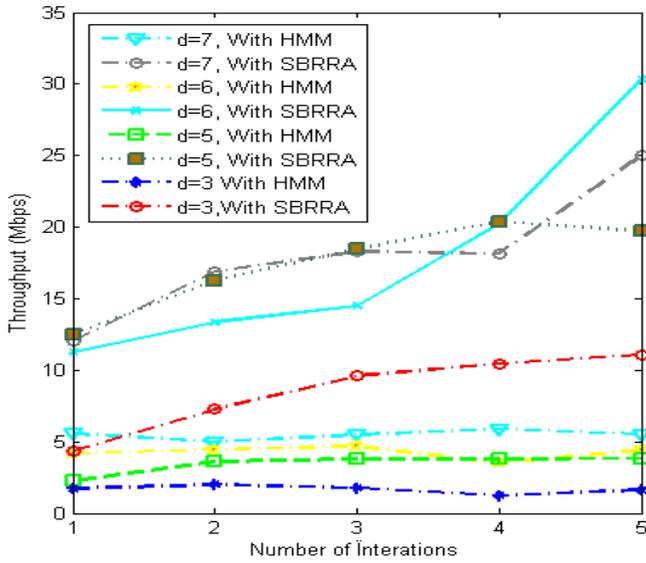

Fig. 15. Throughput v/s number of iterations for n=3, 5, 6 and 7; comparison between HMM and SBRRA

### C. For high User density (N=100)

Similar to the analysis for the above two cases, for low and medium user densities, the obtained SINR, throughput and MOS for an ultra-dense network, with N=100 is studied, i.e high user density. For a higher value of N, the number of pairs formed are also higher. Here, $d$ varies from 3 to 24. Taking $d$=3, 7 and 10, performance indicators for N=100 are depicted in Table XII. SINR values are less at some iterations, due to concentrated density of pairs. Even with a large number of pairs, throughput and MOS values obtained are high, effectively meeting the QoS and QoE demands of the D2D pairs.

TABLE XII
PERFORMANCE METRICS FOR d=3, 7, 10, for N=100

| Parameter | Iterations→ | | 1 | 2 | 3 | 4 | 5 |
|---|---|---|---|---|---|---|---|
| SINR (dB) | $d$=3 | HMM | 51 | 52 | 52 | 72 | 105 |
| | | SBRRA | 50 | 78 | 85 | 71 | 72 |
| | $d$=7 | HMM | 192 | 255 | 291 | 208 | 257 |
| | | SBRRA | 202 | 89 | 268 | 241 | 275 |
| | $d$=10 | HMM | 240 | 180 | 270 | 310 | 260 |
| | | SBRRA | 272 | 201 | 302 | 341 | 282 |
| Throughput (Mbps) | $d$=3 | HMM | 3.06 | 3.12 | 3.12 | 4.3 | 6.27 |
| | | SBRRA | 7.13 | 10.6 | 18.4 | 21 | 23.05 |
| | $d$=7 | HMM | 11.5 | 15.2 | 17 | 12 | 15 |
| | | SBRRA | 18.5 | 40 | 41 | 54.7 | 54.9 |
| | $d$=10 | HMM | 15 | 22.6 | 19.5 | 19 | 14.9 |
| | | SBRRA | 47 | 48 | 49.8 | 64.5 | 75.6 |
| MOS | $d$=3 | HMM | 4.4 | 4.4 | 4.5 | 4.9 | 4.94 |
| | | SBRRA | 4.8 | 4.34 | 4.75 | 4.9 | 4.98 |
| | $d$=7 | HMM | 4 | 4.3 | 4.9 | 4.9 | 4.8 |
| | | SBRRA | 4.9 | 4.9 | 4.8 | 4.75 | 4.9 |
| | $d$=10 | HMM | 4.53 | 4.48 | 4.5 | 4.2 | 4.08 |
| | | SBRRA | 4.7 | 4.28 | 4.62 | 4.6 | 4.9 |

### D. Complexity Analysis

The usefulness of the SBRRA scheme can be clearly validated by the simulation results, for a 5G scenario, with variable user densities. The proposed scheme focuses on use of three-sector antenna at the base station for effective offloading, and enhancing system performance. The impact of all the interfering factors on a D2D pair is depicted in Eq. (2). Since the aim of this paper to evaluate the performance of a cellular network using sector antennas at the BS, the impact of interference power on the cellular network, with and without the use of sector approach is studied, for analyzing the complexity of the network with D2D communication. It is assumed that for both the cases (with and without SBRRA), there are equal number of pairs and all are demanding the same application (A3 assumed here, for complexity analysis). The interference power rises with increasing number of pairs, with pairs formed adjacent to each other raising the co-tier interference. As can be seen from Fig. 16., the complexity of the network, is less, when using SBRRA, and more without sector approach. This depicts the efficiency of the use of sector antennas, in terms of reduced interference for ongoing D2D communication. For pair 7, it is observed that complexity of SBRRA is higher than the scheme not employing SBRRA. This can be related to low SINR for concentrated pair density, as depicted in Fig. 12, and Table IX.

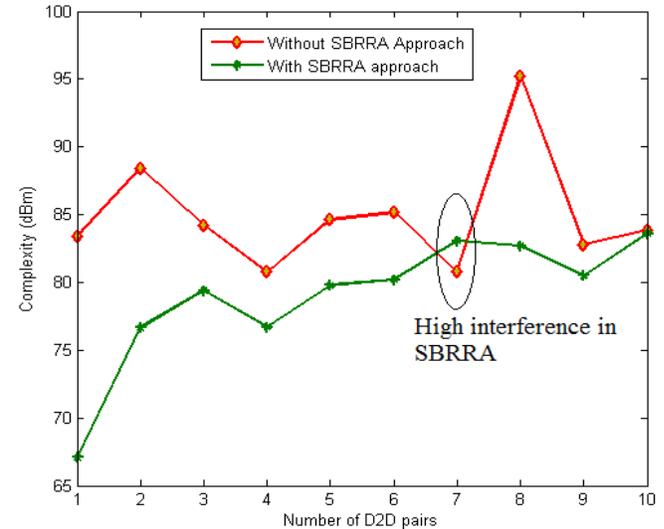

Fig. 16. Complexity v/s Number of D2D pairs

### VII. CONCLUSION

Adaptive RB allocation in D2D communication in 5G WCN has been addressed in this paper, for warranted QoS and QoE. Use of highly directional antennas, dividing the cell into three sectors, is facilitated at the BS, for improved system performance and capacity, with the provision of significant offloading at the BS. The effectiveness of the use of sector approach has been illustrated by the simulation results. For a random number of users in a single sector, D2D pair formation takes place on the basis of the distance constraint. After pair formation, the suitable cellular reuse candidate (RB sharing partner) for each D2D pair is selected. Then optimal number of RBs are allocated to the pairs, depending upon the demanded application (A1, A2 or A3). The performance indicators for QoS and QoE are throughput, and MOS, respectively. Higher throughput and MOS are achieved with SBRRA, due to RB reuse from the same cellular user in successive iterations, which is proposed by SBRRA. The values are compared with the results obtained by RB allocation using HMM also. The sector approach helps in reducing interference in the cellular network, thus reducing complexity and improving the system performance.

This proposed SBRRA technique can be extended for further research in 5G. Various other applications can be tested, apart from those which have been considered in this paper. Power levels at the base station can be optimized with this approach, thus enabling higher energy efficiency in the 5G WCNs.

## APPENDIX

*Lemma 1:* A cellular user providing favorable channel conditions to different pairs during different iterations result in boosting of the overall network throughput, with the resultant throughput given as a summation of throughputs in different iterations.

*Proof:* Let there be pairs $j$ and $(j+1)$ formed during Iteration $n$, $\forall j, (j+1) \in \check{D}$ and $n \in \mathbb{Q}$. When $i^{th}$ cellular user shares '$k$' RBs with $j^{th}$ D2D pair and $(i+1)^{th}$ cellular user shares $k'$ RBs with $j+1^{th}$ pair, $i, (i+1) \in \mathbb{C}$ and $k, k' \in \mathbb{R}$, the count of shared RBs for each being application dependent, then throughput for $j^{th}$ pair is

$$T_j^i(n) = \beta \sum_{k \in \mathbb{R}} \log_2(1 + SINR_j^{k,i}) \quad (20)$$

And for $j+1^{th}$ pair is

$$T_{j+1}^{i+1}(n) = \beta \sum_{k' \in \mathbb{R}} \log_2\left(1 + SINR_{j+1}^{k',i+1}\right) \quad (21)$$

Thus, total system throughput during the $n^{th}$ iteration is given by

$$T_d(n) = T_j^i(n) + T_{j+1}^{i+1}(n) \quad (22)$$

Let us assume that during Iteration $n+1$, again there are two pairs: $j$ and $j+1$. Let us assume that the $i^{th}$ cellular user is providing highest channel gain to pair $j$, again in this iteration, and has sufficient count of RBs for sharing. When $i^{th}$ cellular user shares $k''$ RBs with $j^{th}$ pair, the throughput achieved by the $j^{th}$ pair is given by

$$T_j^i(n+1) = \beta \sum_{k'' \in \mathbb{R}} \log_2\left(1 + SINR_j^{k'',i}\right) \quad (23)$$

However, the $i^{th}$ cellular user was involved in RB sharing during the $n^{th}$ iteration as well. Its participation in RB sharing for the next iteration results in total throughput for $j^{th}$ pair, in iteration $n+1$ to be expressed, as per SBRRA, as

$$T_j^i(n+1) = \beta \sum_{k'' \in \mathbb{R}} \log_2\left(1 + SINR_j^{k'',i}\right) + T_j^i(n) \quad (24)$$

Therefore, using the proposed approach, the throughput can be substantially enhanced. This process continues till $i^{th}$ cellular user has sufficient count of RBs and is providing favorable channel states for RB sharing to the pairs formed in different iterations.

In general, when a cellular user '$i$' is involved in RB sharing during multiple iterations, with $l$ number of D2D pairs, then the resultant throughput for $l$ pairs, for $n$ number of iterations, such that $l \in \check{D}$, is given by

$$T_l^i = \sum_{n \in \mathbb{Q}} \sum_{p=1}^{l} \sum_{k \in \mathbb{R}} \log_2(1 + SINR_p^{k,i}) \quad (25) \blacksquare$$

**ACKNOWLEDGEMENT:** The authors gratefully acknowledge the support provided by 5G and IoT Lab, DoECE, and TBIC, Shri Mata Vaishno Devi University, Katra, Jammu.

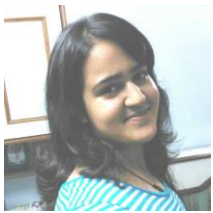

**Miss. Pimmy Gandotra (S'16)** received the B.E. degree in Electronics and Communication Engineering from Jammu University, Jammu and Kashmir, India, in 2013 and the M.Tech degree in Electronics and Communication Engineering from Shri Mata Vaishno Devi University in 2016. She is currently pursuing the Ph. D degree in Electronics and Communication Engineering at Shri Mata Vaishno Devi University, Katra, Jammu and Kashmir, India

Her research interest includes the emerging technologies of 5G wireless communication network and currently doing her research work on Power Optimization in next generation networks.

She has received teaching assistantship from MHRD from 2014-2016. She is a student member of Institute of Electrical and Electronics Engineers (IEEE).

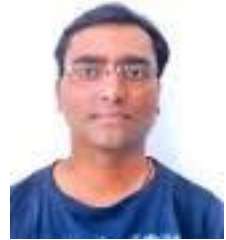

**Dr. Rakesh K Jha (S'10, M'13)** is currently an Assistant Professor in School of Electronics and Communication Engineering, Shri Mata Vaishno Devi University, katra, Jammu and Kashmir, India. He is carrying out his research in wireless communication, power optimizations, wireless security issues and optical communications. He has done B.Tech in Electronics and Communication Engineering from Bhopal, India and M.Tech from NIT Jalandhar, India. Received his PhD degree from NIT Surat, India in 2013.

He has published more than 30 International Journal papers and more than 20 International Conference papers. His area of interest is Wireless communication, Optical Fiber Communication, Computer networks, and Security issues.

Dr. Jha's one concept related to router of Wireless Communication has been accepted by ITU (International Telecommunication Union) in 2010. He has received young scientist author award by ITU in Dec 2010. He has received APAN fellowship in 2011 and 2012, and student travel grant from COMSNET 2012. He is a senior member of IEEE, GISFI and SIAM, International Association of Engineers (IAENG) and ACCS (Advance Computing and Communication Society).

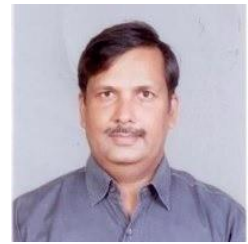

**PROF. SANJEEV JAIN**, born at Vidisha in Madhya Pradesh in 1967, obtained his Post Graduate Degree in Computer Science and Engineering from Indian Institute of Technology, Delhi, in 1992. He later received his Doctorate Degree in Computer Science & Engineering and has over 24 years' experience in teaching and research. He has served as Director, Madhav Institute of Technology and Science (MITS), Gwalior. Presently, he is working as a vice chancellor at Shri Mata Vaishno Devi University, Katra.

Besides teaching at Post Graduate level Professor Jain has the credit of making significant contribution to R & D in the area of Image Processing and Mobile Adhoc Network. He has guided Ph.D. Scholars and has undertaken a number of major R & D projects sponsored by the Government and Private Agencies. His work on Digital Watermarking for Image Authentication is highly valued in the research field. He is also a member of Associaltion for Computing Machinery (ACM).